\documentclass[journal]{IEEEtran}
\IEEEoverridecommandlockouts
\usepackage{amsmath, amssymb, amsfonts}

\usepackage{graphicx}
\usepackage{xcolor}
\usepackage[table]{xcolor}
\usepackage{booktabs} 
\usepackage{wrapfig}
\usepackage{comment}
\usepackage{subcaption}

\usepackage[normalem]{ulem}
\usepackage{subcaption}

\usepackage{multicol}
\usepackage{color}
\usepackage{siunitx}

\usepackage{soul}
\usepackage{dblfloatfix}
\usepackage{epstopdf}

\usepackage{comment}
\usepackage[numbers, sort&compress, square]{natbib}
\usepackage{enumerate}
\usepackage{mathtools}
\usepackage{amsmath}

\usepackage{siunitx}
\usepackage{multirow}
\usepackage{threeparttable} 
\usepackage[noend]{algpseudocode}
\usepackage{pdfpages}
\usepackage{float}
\usepackage{scalerel}
\usepackage{tikz}
\usepackage{amssymb} 
\usepackage{enumitem}
\usepackage{makecell}

\usepackage{hyperref}

\usepackage{cases}

\usepackage{multicol,lipsum,microtype}
\usepackage{marvosym}

\usepackage{amsmath}

\newcommand{\p}{\pmb{\p}}

\usepackage[intoc]{nomencl}

\makenomenclature
\renewcommand{\nomgroup}[1]{%
    \ifthenelse{\equal{#1}{A}}{\item[\textit{Indices and Sets}] }{%
    \ifthenelse{\equal{#1}{B}}{\vspace{20pt} \item[\textit{Binary Variables}]}{}}{%
    \ifthenelse{\equal{#1}{C}}{\vspace{20pt} \item[\textit{Continuous Positive Variables}]}{}}{%
    \ifthenelse{\equal{#1}{D}}{\vspace{20pt} \item[\textit{Parameters}]}{}}}
     
\usepackage{titlesec}

\titlespacing*{\section}{0pt}{4pt}{2pt}
\titlespacing*{\subsection}{0pt}{4pt}{1pt}
\titlespacing*{\subsubsection}{0pt}{3pt}{1pt}

\usepackage{booktabs}
\usetikzlibrary{svg.path}
\usepackage[ruled,vlined]{algorithm2e}
\SetKwInput{KwInput}{Input}                
\SetKwInput{KwOutput}{Output}              
\usepackage{tikz}
\usepackage{pgfplots}
\usepackage{amsfonts} 
\pgfplotsset{compat=1.18}

\definecolor{orcidlogocol}{HTML}{A6CE39}
\tikzset{
orcidlogo/.pic={
\fill[orcidlogocol] svg{M256, 128c0, 70.7-57.3, 128-128, 128C57.3, 256, 0, 198.7, 0, 128C0, 57.3, 57.3, 0, 128, 0C198.7, 0, 256, 57.3, 256, 128z};
\fill[white] svg{M86.3, 186.2H70.9V79.1h15.4v48.4V186.2z}
         svg{M108.9, 79.1h41.6c39.6, 0, 57, 28.3, 57, 53.6c0, 27.5-21.5, 53.6-56.8, 53.6h-41.8V79.1z M124.3, 172.4h24.5c34.9, 0, 42.9-26.5, 42.9-39.7c0-21.5-13.7-39.7-43.7-39.7h-23.7V172.4z}
         svg{M88.7, 56.8c0, 5.5-4.5, 10.1-10.1, 10.1c-5.6, 0-10.1-4.6-10.1-10.1c0-5.6, 4.5-10.1, 10.1-10.1C84.2, 46.7, 88.7, 51.3, 88.7, 56.8z};
}
}

\newcommand\orcidicon[1]{\href{https://orcid.org/#1}{\mbox{\scalerel*{
\begin{tikzpicture}[yscale=-1, transform shape]
\pic{orcidlogo};
\end{tikzpicture}
}{|}}}}

\def\BibTeX{{\rm B\kern-.05em{\sc i\kern-.025em b}\kern-.08em
T\kern-.1667em\lower.7ex\hbox{E}\kern-.125emX}}

\SetKwInput{KwInput}{Input}                
\SetKwInput{KwOutput}{Initialization}      

\begin{document}
\bstctlcite{IEEEexample:BSTcontrol}

\twocolumn
\title{Scarcity Pricing for State-of-Charge Management
in Real-Time Markets Under Intra-Day Uncertainty

\thanks{This work was funded in part by the U.S. Dept. of Energy, ARPA-E \# DE-AR0001281,  DOE Award \# GD0000958  and PJM Interconnection. Computations for this research were performed on the Pennsylvania State University’s Institute for Computational and Data Sciences’ Roar supercomputer.}

}

\author{Shailesh Wasti, \IEEEmembership{Student Member, IEEE}, Aravind Retna Kumar, \IEEEmembership{Student Member, IEEE}, \\ Sushant Varghese, \IEEEmembership{Member, IEEE}, Anthony Giacomoni, \IEEEmembership{Senior Member, IEEE},
Parviz Alivand, \IEEEmembership{Member, IEEE}, \\ Yujie Hu, \IEEEmembership{Member, IEEE}, and 
Mort Webster, \IEEEmembership{Member, IEEE}
\thanks{ Shailesh Wasti, Aravind Retna Kumar, and Mort Webster are with the Department of Energy and Mineral Engineering, The Pennsylvania State University, State College, PA 16801, USA (e-mails: sjw6236@psu.edu, aravind.kumar@psu.edu, mdw18@psu.edu). Sushant Varghese was with the Department of Industrial and Manufacturing Engineering, The Pennsylvania State University, State College, PA 16801, USA. Anthony Giacomoni is with the Advanced Analytics Department, PJM Interconnection, Audubon, PA 19403, USA (e-mail: anthony.giacomoni@pjm.com). Parviz Alivand and Yujie Hu are with the Market Development, ISO New England  Inc., Holyoke, MA 01040, USA (e-mails: palivand@iso-ne.com,yhu@iso-ne.com.)}
}
\IEEEaftertitletext{\vspace{-2.5em}}
\maketitle
\begin{abstract}
The growth in renewable resources will increase net load variability and forecast error.  Energy storage resources can provide flexibility when ramp capability is limited and contribute to load leveling over the day.  The challenge for system operators is how to balance  these two functions in real-time, when updated net load forecasts are available for a limited lookahead horizon. U.S. electricity markets are seeking ways to manage state-of-charge (SoC) within their energy markets to more efficiently utilize these resources. We propose a method to manage energy storage SoC in a real-time market by extending Hogan’s (2013) scarcity pricing framework.  A day-ahead SoC schedule is modified using existing multi-interval real-time unit commitment engines as net load forecasts change.  To address 30-minute ahead forecast errors, an updated 5-minute SoC target and a dynamically updated shortage price for violating the constraint are provided to each economic dispatch.  We simulate PJM’s energy markets for 365 days and demonstrate reliability improvements relative to a strategy that enforces the day-ahead schedule. We show cost reductions and improved price formation relative to a strategy that updates target quantities but uses an administratively chosen shortage penalty. 
\end{abstract}


\nomenclature[A,01]{$g \in G$}{Set of generators}
\nomenclature[A,02]{$s \in \mathcal{S}$}{Set of battery energy storage resources}
\nomenclature[A,03]{$t \in T$}{Set of time periods}
\nomenclature[A,04]{$j \in J$}{Set of points in the operating reserve demand curve}

\nomenclature[B,01]{$u_{g,t}$}{Commitment status of generator $g$ at time $t$}
\nomenclature[B,02]{$\lambda_{g,t,\sigma}$}{Startup indicator of type $\sigma$ for generator $g$ at time $t$}
\nomenclature[B,03]{$u_{s,t}^{ch}$}{Charging status of storage $s$ at time $t$}
\nomenclature[B,04]{$u_{s,t}^{dis}$}{Discharging status of storage $s$ at time $t$}

\nomenclature[C,01]{$w_{g,t}$}{Power above minimum load of generator $g$ at time $t$ (MW)}
\nomenclature[C,02]{$f_{g,t}$}{Fuel cost of generator $g$ at time $t$ (\$)}

\nomenclature[C,03]{$e_{s,t}$}{Energy level of energy storage $s$ at time $t$ (MWh)}
\nomenclature[C,04]{$p_{s,t}^{ch}$}{Charging rate of storage $s$ at time $t$ (MW)}
\nomenclature[C,05]{$p_{s,t}^{dis}$}{Discharging rate of storage $s$ at time $t$ (MW)}

\nomenclature[C,06]{$r^{\text{sr}}_{g,t}$}{Synchronized reserve from generator $g$ at time $t$ (MW)}
\nomenclature[C,07]{$r^{\text{sr-or}}_{g,t}$}{Synchronized reserve component of operating reserve from $g$ at $t$ (MW)}
\nomenclature[C,08]{$r^{\text{nsr-pr}}_{g,t}$}{Non-spinning reserve component of primary reserve from $g$ at $t$ (MW)}
\nomenclature[C,09]{$r^{\text{nsr-or}}_{g,t}$}{Non-spinning reserve component of operating reserve from $g$ at $t$ (MW)}
\nomenclature[C,10]{$p_{s,t}^{sr}$}{Synchronized reserve allocation of storage $s$ at time $t$ (MW)}

\nomenclature[C,11]{$\psi^{\text{ul}}_t$}{Unserved load at time $t$ (MW)}
\nomenclature[C,12]{$\psi^{\text{sr}}_{j,t}$}{Unmet synchronized reserve at level $j$ and time $t$ (MW)}
\nomenclature[C,13]{$\psi^{\text{pr}}_{j,t}$}{Unmet primary reserve at level $j$ and time $t$ (MW)}
\nomenclature[C,14]{$\psi^{\text{or}}_{j,t}$}{Unmet operating reserve at level $j$ and time $t$ (MW)}
\nomenclature[C,15]{$\psi_s^{ue}$}{Unmet SoC target of storage $s$ at time $t$ (MW)}

\nomenclature[D,01]{$\Delta \tau$}{Time resolution of the model (hrs.)}
\nomenclature[D,02]{${C}_{g,\sigma}^{\text{start}}$}{Startup cost of generator $g$ of type $\sigma$ (\$)}
\nomenclature[D,03]{$\overline{P}_{g}$}{Maximum power output of generator $g$ (MW)}
\nomenclature[D,04]{$\underline{P}_{g}$}{Minimum power output of generator $g$ (MW)}
\nomenclature[D,05]{${C}_{g}^{\text{vom}}$}{Variable O\&M cost of generator $g$ (\$/MWh)}
\nomenclature[D,05]{${C}_{s}^{\text{vom}}$}{Variable O\&M cost of ESR $s$ (\$/MWh)}
\nomenclature[D,06]{${C}^{\text{ul}}$}{Penalty cost for unserved load (\$/MWh)}
\nomenclature[D,07]{${C}^{\text{sr}}_j$}{Penalty cost for unmet synchronized reserve at level $j$ (\$/MWh)}
\nomenclature[D,08]{${C}^{\text{pr}}_j$}{Penalty cost for unmet primary reserve at level $j$ (\$/MWh)}
\nomenclature[D,08]{${C}^{\text{ue}}_j$}{Penalty cost for SoC target violation (\$/MWh)}
\nomenclature[D,09]{$D_t$}{Net load at time $t$ (MW)}
\nomenclature[D,10]{$R^{\text{sr}}_{j}$}{Synchronized reserve requirement at level $j$ (MW)}
\nomenclature[D,11]{$R^{\text{or}}_{j}$}{Operating reserve requirement at level $j$ (MW)}
\nomenclature[D,11]{$R^{\text{pr}}_{j}$}{Primary reserve requirement at level $j$ (MW)}
\nomenclature[D,12]{$\overline{E}_{s,t}$}{Maximum energy level of storage $s$ at time $t$ (MWh)}
\nomenclature[D,13]{$\underline{E}_{s,t}$}{Minimum energy level of storage $s$ at time $t$ (MWh)}
\nomenclature[D,14]{$\overline{P}_{s}$}{Maximum power rating of storage $s$ (MW)}
\nomenclature[D,15]{$\underline{P}_{s}$}{Minimum power rating of storage $s$ (MW)}
\nomenclature[D,16]{$\eta^{ch}$}{Charging efficiency of storage}
\nomenclature[D,17]{$\eta^{dis}$}{Discharging efficiency of storage}
\nomenclature[D,18]{$E^0$}{Initial and terminal SoC as a percentage of max capacity}
\nomenclature[D,19]{$\mathcal{T}$}{Reserve requirement duration for revenue payment}
\nomenclature[D,20]{$\alpha$}{Fraction of energy capacity for initial/final SoC constraint}

\section{Introduction}
The growing capacity of variable renewable energy sources will increase the variability of net load (electricity demand minus renewable generation), exacerbate intra-day and intra-hour forecast errors, and result in steeper ramps during afternoon and evening hours \cite{Rand2024QueuedUp,hong_toward_2026, singhal2024navigating, Yurdakul2025}. Energy Storage Resources (ESRs) have the potential to reduce system costs via load-shifting and to maintain reliability by providing a flexible, responsive energy source at times of system scarcity \cite{ armstrong2022futureof,sioshansi2021energy}.  

FERC Order 841 \cite{FERC841} mandated that Independent System Operators (ISOs) and Regional Transmission Organizations (RTOs) develop market rules to facilitate the participation of ESRs in energy, ancillary services, and capacity markets. All North American market operators  allow ESR participation.  In most markets, ESRs self-schedule, and the market operator does not manage the State-of-Charge (SoC) \cite{singhal2024navigating}. Only CAISO and NYISO actively manage SoC in their markets. PJM optimizes only pumped storage hydro in its Day-Ahead (DA) Market.

There is growing evidence that ESR participation without management of SoC in the Real-Time Market (RTM) may not provide the expected benefits and could lead to undesirable market outcomes.  In 2025, out-of-market reliability actions in ERCOT's RTM increased by 176\% from the previous year, and make-whole payments exceeded \$21 million.  The Independent Market Monitor attributes these outcomes, in part, to the strict enforcement of the DA SoC schedule instead of re-optimizing in their real-time Reliability Unit Commitment \cite{potomac2025ercot}. CAISO, which introduced end-of-hour SoC constraints in 2021 and reoptimizes storage within its multi-interval real-time dispatch, notes the persistence of suboptimal charging/discharging behavior and is considering market design changes to address the issue. PJM, ISO-NE, and others are actively exploring market design changes to manage real-time SoC \cite{caiso2025special,parakkuth2024ERCOT,pjm2026enhancing}. 

The need for market operators to optimize SoC is supported by economic and optimization theory. Jiang and Sioshansi \cite{ jiang2023duality} argue on the basis of duality theory that market operators should optimize the aggregate state-of-charge. Baldick  \cite{baldick2026} argues that markets should manage SoC to reflect the physical characteristics of energy-limited resources and maximize market efficiency. Similarly, Munoz and co-authors \cite{ munoz2018economic} show that when opportunity cost diverges from the marginal cost of energy that a pure bid-based (non-optimized storage) dispatch will not be economically efficient.

Most ESR studies have focused on optimal bidding strategies from the resource perspective \cite{bahramara_modeling_2022,zheng2023energy}. The benefits to consumers from optimizing ESRs over longer horizons (e.g., 24 hours) is well established \cite{sakti2018review, robertson2025chargingup, armstrong2022futureof}. Many studies of optimal allocation of storage apply deterministic or stochastic DA unit commitment models \cite{lu_multi-stage_2020,tang2020reserve,lorca_multistage_2017}. 

Even an SoC schedule that is optimal within the DA market requires real-time (RT) adjustment to compensate for DA net load forecast errors, which can be quite large. In 2023, 10\% of PJM's DA forecasts were more than 2.8 GW too high and 10\% were more than 3.1 GW too low \cite{pjm2026marketsreport,pjm2026systemsoperations}. CAISO and MISO have also experienced DA forecast errors of 3 to 4 GW \cite{spyrou2025flexibility}. Some studies have explicitly explored mechanisms for RT adjustments to DA SoC schedules to address DA forecast errors \cite{zhao2019multi,goyal2024design}, and others have implicitly addressed the issue with proposed stochastic DA model frameworks \cite{lu_multi-stage_2020,tang2020reserve,lorca_multistage_2017}.

In addition to DA forecast errors, RT net load forecast errors can also be quite large at times. RT clearing engines, including real-time unit commitment (RT-UC) and real-time economic dispatch (RT-ED), use rolling forecasts that are updated between each market clearing and can have significant errors \cite{ghadimi2024stochastic, Chen2023}.  Powell and coauthors \cite{simao2017challenge, ghadimi2024stochastic} have argued for the importance of addressing these intra-RT forecast errors by improving SoC management in RT. PJM's 60-minute ahead forecasts for 2022-2023 were more than 800 MW too low 10\% of the time and more than 800 MW too high 10\% of the time. 30-minute ahead forecasts were more than 600 MW too low 10\% of the time and more than 560 MW too high 10\% of the time. Moreover, the direction of RT forecast errors often persist for up to several hours, leading to a cumulative ramp capacity shortfall \cite{pjm2026marketsreport, pjm2026systemsoperations}. 

When net load exceeds the most recent forecast, the additional energy must be provided by the system's ramping capability.  If that is insufficient, operators may resort to out-of-market commitments. Increasing frequency of these manual interventions has been noted in several markets, including CAISO \cite{CAISO2022Performance} and MISO \cite{Chen2023}.  In 2025, PJM operators made 5855 manual commitments for a total of 750 GWh over the year, an increase of 27\% and 69\%, respectively from two years prior \cite{PJMDataMiner}. An alternative is to use energy from ESRs that was scheduled for later in the day and replenish it afterwards.  However, using stored energy for unexpected ramping needs requires deviating from the DA SoC schedule.

Consider the perspective of the system operator at one moment in the middle of one day.  Assume the ESR charging/discharging schedule was set in the DA market clearing based on the DA load forecast. In RT, the DA SoC target is downscaled from hourly to 5-minute targets to define the minimum energy level for that interval. Before the next RT-UC, assume the updated net load forecast has deviated from prior forecasts and now indicates a ramp capability shortage during the next 30 minutes. The problem we address is how the operator can decide whether to deviate from the DA schedule and use stored energy for unanticipated ramp shortages over the next half-hour or, alternatively, to commit additional thermal capacity. The horizons for RT clearing engines are too short to allow explicit re-optimization to the end of the day.

Many markets enforce or are considering an end-of-hour SoC constraint in RT markets, which can be taken from the DA schedule or a re-optimized schedule in real-time. SoC management in RT requires a method for setting two quantities: 1) The adjusted SoC target for the next interval, and 2) The penalty for violating the SoC target. The SoC schedule can be adjusted within  the RT-UC, which has better information on the current system state and the net load forecast than was used in the DA.  Deviations from the updated SoC target may be required to address a revised load forecast since the prior RT-UC. If the realized load is higher than forecast, the penalty for a SoC shortage provides information about the system's opportunity cost of using one additional MWh beyond the target. Setting a low value for this penalty will lead to more frequent discharging and less energy available later in the day.  Conversely, a high penalty will discourage deviating from the DA schedule, even if it is a lower cost solution. 

We propose a method for managing the aggregate SoC in Real-Time Markets that re-optimizes throughout the day as forecasts change.  The initial SoC schedule is determined in the Day-Ahead Market. In real-time, SoC targets are adjusted using the existing multi-interval RT-UC, which enables implementation in markets with single-interval RT-ED engines.  The modified SoC target from the RT-UC solution is provided to RT-ED along with a penalty for not meeting that target.  

The penalty for a SoC shortage is dynamically determined and provides a forward-looking signal that reflects the system's marginal opportunity cost of depleting stored energy below the target. We extend Hogan's \cite{hogan2013electricity} formulation for the marginal value of reserves to manage SoC. The marginal value of reserves is calculated \textit{ex ante} before each RT-ED using the historical forecast error and the current system state.

Our approach is distinct from existing approaches in three respects. First, the value of ramping capability is based on the aggregate ramping capability from all resources, including thermal generators and ESRs.  The marginal value of surplus ramping capability at each RTM clearing provides the penalty for deviating from the SoC target.
Second, we integrate the proposed mechanism into the sequential market-clearing process used by RTOs/ISOs, facilitating implementation with minimal structural changes to existing market architectures and using information available during real-time. Third, the marginal value of SoC shortage is dynamically calculated \textit{ex ante} for better dispatch decisions and price formation. 

We compare our proposed SoC management approach to a strategy that enforces the DA SoC schedule and to a strategy that adjusts the SoC targets but administratively sets a constant shortage penalty. Strategies are compared for a simulation of PJM’s energy markets, which solves the DA market and then solves a rolling horizon sequence of RT-UC every 15 minutes (two-hour horizon) and three 5-minute RT-ED clearings, and represents forecast errors that change between RT-UC executions.  The rolling horizon market simulation framework is similar to several other studies \cite{ela2012studying,simao2017challenge,cartuyvels2022calibration}. We simulate 365 days, using historical 5-minute forecast and realized load, wind, and solar, and historical forecast errors throughout each day.  We demonstrate that when SoC is not reoptimized in real-time, instead simply enforcing the DA hourly SoC schedule, we observe  reliability issues, cost impacts and out-of-market commitments similar to those observed in ERCOT and CAISO. We then demonstrate the improvements in market outcomes from our proposed strategy for the same 365 days, with the same rolling forecast errors.
\begin{figure*}
    \centering    \includegraphics[width=0.90\linewidth]{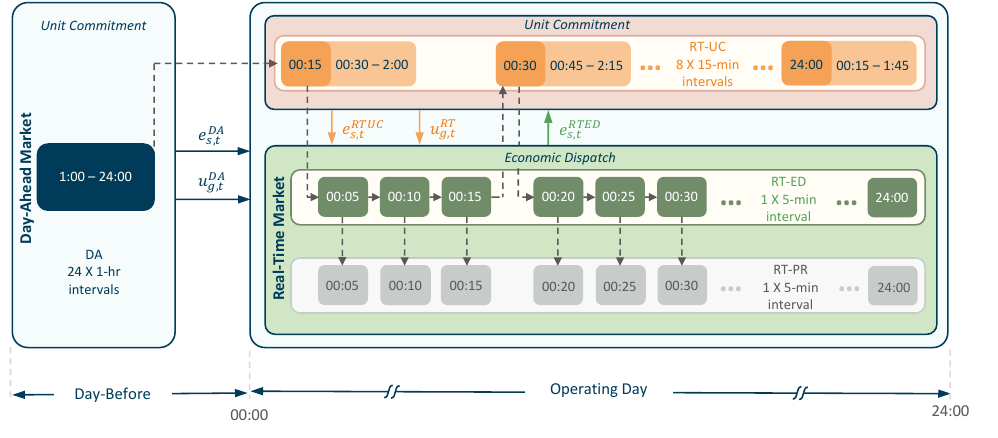}
    \caption{Electricity Market Simulation Framework}
    \label{fig:schematic}
    \vspace{-1.5em}
\end{figure*}

Our contributions are:
\begin{enumerate}
    \item A method for updating  DA SoC targets in RT using RT-UC, requiring minimal changes to operational structures.
    \item A practical approach to managing SoC in single-interval RT-ED that uses an updated SoC target and allows a SoC shortage at a dynamically adjusted penalty.  The penalty is set to the marginal value of reserves \cite{hogan2013electricity}.
    \item A simulation of PJM’s energy markets for 365 days that demonstrates improvements in cost, reliability, and price formation under this approach relative to strategies that enforce the DA SoC schedule or that update the SoC targets but set SoC shortage penalties administratively.
    \end{enumerate}

The remainder of the paper is organized as follows.  Section \ref{sec:methodology} presents the formulations that represent energy storage in the market simulation, and the implementation of alternative SoC management strategies. Section \ref{sec:ResultsIllustrative} illustrates the concepts using a simple example. Section \ref{sec:ResultsPJM} compares market outcomes across RT SoC management strategies using a simulation of PJM's energy markets. Section \ref{sec:Discussion} discusses the implications of market designs for energy storage management.

\section{Methodology} \label{sec:methodology} 
This section presents the market simulation framework used to reproduce the key features of wholesale energy markets. The Electricity Market Simulation Framework (EMSF) ~\cite{EMSF_2025} includes a DA commitment and dispatch model and a rolling horizon representation of the RT market that iteratively solves an RT-UC model and a sequence of single-interval RT-ED models.  The EMSF represents net load forecast errors in the DA and RT that evolve stochastically over each simulated operating day. The EMSF framework is similar to the rolling-horizon simulation frameworks SMART-ISO~\cite{simao2017challenge}, FESTIV~\cite{ela2012studying}, and that of Cartuyvels and Papavasiliou~\cite{cartuyvels2022calibration}. Fig. \ref{fig:schematic} illustrates the temporal structure and information flow of the framework.

\subsection{Model Formulation}
\label{sec:formulation}
We have documented elsewhere the complete formulations for the DA, RT-UC, RT-ED, and Real-Time Pricing (RT-PR) models \cite{EMSF_2025}. For brevity, we present here only the modifications to the formulation necessary to represent SoC management. For each component model, we present the modifications for the reference version of SoC management, which strictly enforces the DA SoC targets in the real-time engines. We then present the formulation for the alternative SoC management strategies.

\subsubsection{Changes to the Day-Ahead Model} 
The DA model is formulated as a mixed-integer linear program (MILP) and solves the least-cost commitment and dispatch for a 24-hour horizon with hourly time steps, based on a day-ahead net load forecast. The resulting commitment schedule for slow-start units is fixed in the subsequent real-time engines. The DA model is extended here to solve for a schedule of hourly SoC targets that meets a terminal (end-of-day) condition. The DA SoC schedule constrains ESR discharging in the real-time clearing engines, which operate over shorter horizons. 

\emph{Objective Function:} The objective function \eqref{eq:obj_DA} minimizes the total cost of providing energy and ancillary services, including penalties for unserved load and reserve shortages. The variables are the power output $p_{g,t}$ of each thermal generator $g$ at interval $t$, discharging $p^{\text{dis}}_{s,t}$ and charging $p^{\text{ch}}_{s,t}$ for energy storage unit $s$ at time $t$, and generator startup decisions $\upsilon_{g,t}$. $C_g^{\text{OP}}$ is the operating cost (fuel and operation and maintenance (O\&M)) of generator $g$, $C_{g}^{\text{start}}$ is the startup cost, and $C^{\text{OP}}_{s}$ the O\&M cost of storage unit $S$. Unserved load $\psi^{\text{ul}}_t$ and shortages of synchronized reserves $\psi^{\text{sr}}_{j,t}$, primary reserves $\psi^{\text{pr}}_{j,t}$, and operating reserves $\psi^{\text{or}}_{j,t}$ incur penalties $C^{\text{ul}}$, $C^{\text{sr}}_j$, $C^{\text{pr}}_j$, and $C^{\text{or}}_j$, respectively, for the $j^{\text{th}}$ block of each reserve's demand curve.
\begin{align}
\begin{split}
    \min & \underbrace{\sum_{g,t} C_{g}^{\text{start}} \upsilon_{g,t}}_{\text{start-up cost}} + 
    \underbrace{\sum_{g,t} C_g^{\text{OP}} p_{g,t}}_{\text{generation cost}} + \underbrace{\sum_{s,t} C^{\text{OP}}_s (p^{\text{dis}}_{s,t} + p^{\text{ch}}_{s,t})}_{\text{ESR cost}} 
    + \\ 
    &  \underbrace{\sum_t C^{\text{ul}} \psi^{\text{ul}}_t}_{\text{cost of unserved load}}  
    + \underbrace{\sum_{j,t} C^{\text{sr}}_j \psi^{\text{sr}}_{j,t} + \sum_{j,t} C^{\text{pr}}_j \psi^{\text{pr}}_{j,t}+ \sum_{j,t} C^{\text{or}}_j \psi^{\text{or}}_{j,t}}_{\text{cost of unmet reserve}} 
\end{split}
\label{eq:obj_DA}
\end{align}
\emph{System Constraints:} Equation \eqref{eq:power_balance} enforces the power balance constraint, ensuring that the net load in each interval, \(D_t\), is met by dispatch from thermal generators and ESR discharge, and unserved load \(\psi^{\text{ul}}_t\). Equations \eqref{eq:sr}, \eqref{eq:primres}, and \eqref{eq:opres} represent synchronized, primary, and operating reserves, respectively. Only online thermal generators and storage can contribute to synchronized reserves.  An offline units can contribute to Primary Reserves or to Operating Reserves depending on its startup time.  In PJM, synchronized and primary reserves have a 10-minute delivery time and operating reserves have a 30-minute delivery time.  In \eqref{eq:sr}, $r^{\text{sr}}_{g,t}$ is the contribution to synchronized reserves from thermal unit $g$, \(r^{\text{sr}}_{s,t}\) is the contribution from ESR \(s\), \(\psi^{\text{sr}}_{j,t}\) is the reserve shortage from the \(j^{\text{th}}\) block of the synchronized reserve demand curve, and the system requirement for block $j$ is $R^{\text{sr}}_j$. Constraint \eqref{eq:primres} enforces primary reserves, which include all synchronized resources and allows offline units \(G^{\text{RTUC-pr}}\) that are eligible to provide primary reserve and can start within the required time-frame. Equation \eqref{eq:opres} extends the set of offline units 
$G^\text{RTUC-or}$ to longer lead-time units. Reserve product prices cascade, and energy storage is eligible to contribute to all three types of reserve.

\begin{align}
     \sum_g p_{g,t} + \sum_{s} p^{\text{dis}}_{s,t} - \sum_{s} p^{\text{ch}}_{s,t} + \psi^{\text{ul}}_t  \geq  D_t & 
    \quad \quad  & \forall t \label{eq:power_balance} 
\end{align}
\begin{align}
     \sum_g r^{\text{sr}}_{g,t} + \sum_{s} r^{\text{sr}}_{s,t} + \sum_j \psi^{\text{sr}}_{j,t} \geq  \sum_j R^{\text{sr}}_j  &  \quad \quad & \forall t \label{eq:sr}
\end{align}

\begin{align}
      \sum_g r^{\text{sr}}_{g,t} + \sum_{g \in G^\text{RTUC-pr}} r^{\text{nsr-pr}}_{g,t} + \sum_{s} r^{\text{sr}}_{s,t} \nonumber +&  \\   \sum_j \psi^{\text{pr}}_{j,t} \geq  \sum_j R^{\text{pr}}_{j,t} &  
     \quad \quad  & \forall t \label{eq:primres} 
\end{align}
\begin{align}
      \sum_g r^{\text{sr-or}}_{g,t} + \sum_{g \in G^\text{RTUC-or}} r^{\text{nsr-or}}_{g,t} + \sum_{s} r^{\text{sr}}_{s,t} \nonumber +&  \\ 
       \sum_j \psi^{\text{or}}_{j,t} \geq  \sum_j R^{\text{or}}_{j,t}&  
     \quad \quad & \forall t \label{eq:opres}
\end{align}

\emph{ESR Physical Constraints:}
Constraints \eqref{eq:energy_limit1} - \eqref{eq: initial_cdn} represent the physical characteristics of energy storage. The energy constraint \eqref{eq:energy_limit1} determines the stored energy level (SoC), $e_{s,t}$, based on the energy level in the previous interval, $e_{s,t-1}$, and the amount of energy charged or discharged. The efficiency of charging and discharging are parameterized, respectively, with \(\eta_s^{\text{dis}}\) and \(\eta_s^{\text{ch}}\). To minimize battery degradation from deep cycles, the total energy in each ESR is kept within a specified minimum, $\underline{E_s}$, and maximum, $\overline{E_s}$, \eqref{eq:energy_limit2}. Constraints \eqref{eq:charge_limit} and \eqref{eq:discharge_limit} ensure that the storage units operate within their maximum power limits. We assume an initial and terminal SoC for each day as a fraction \(\alpha\) of each unit's maximum energy capacity \eqref{eq: initial_cdn}, where \(t_0\) represents the initial time period, and \(t_{|T|}\) denotes the final period, and \(|T|\) is the cardinality of the set \(T\).
\begin{align}
     e_{s,t} = e_{s,t-1} + \eta^{ch} p_{s,t}^{ch}  - \frac{p_{s,t}^{dis}}{\eta^{dis}} \quad \quad \forall s,t \label{eq:energy_limit1}\\
    \underline{E_s} \leq e_{s,t} \leq \overline{E_s} \quad \quad \forall s,t \label{eq:energy_limit2} \\
    0   \leq  p_{s,t}^{ch} \leq \overline{P_s}   \quad \quad \forall s,t \label{eq:charge_limit} \\
 0   \leq  p_{s,t}^{dis} \leq \overline{P_s}   \quad \quad \forall s,t \label{eq:discharge_limit} \\
    e_{s,t} =  \alpha \text{ } \overline{E_s} \quad \quad \forall s,t \in \{t_0,t_{|T|}\}   \label{eq: initial_cdn}
\end{align}

\emph{ESR Reserve Constraints:}
We constrain each ESR's contribution to reserves to ensure deliverability while allowing the model to dynamically allocate SoC between energy and reserves, in contrast with other formulations \cite{padmanabhan_battery_2020}. Our formulation prioritizes energy delivery while allowing for reserve provision within the energy constraints presented above. To account for the energy-limited nature of the ESRs, reserves are allocated only if the ESR can discharge continuously for at least \(\mathcal{T}\) consecutive periods. Constraint \eqref{eq:reserve1} enforces this condition by requiring sufficient available SoC above the minimum threshold, \((e_{s,t} - \underline{E}_s)\), scaled by the discharging efficiency, to deliver the synchronized reserve allocation \(p_{s,t}^{\text{sr}}\) over \(\mathcal{T}\) periods. Constraint \eqref{eq:reserve2} ensures that the total discharging or charging during each interval \(t\) remains within the ESR's maximum power rating. The maximum reserve offer is constrained by the maximum power output of the unit (\(\overline{P}_s\)) less the  scheduled net discharging for that same interval ($p_{s,t}^{\text{dis}} - p_{s,t}^{\text{ch}}$).
\begin{align} 
   r_{s,t}^{\text{sr}} \mathcal{T} &\leq (e_{s,t} - \underline{E}_s)  \eta^{\text{dis}} \quad &&\forall s,t  \label{eq:reserve1} \\
   r_{s,t}^{\text{sr}} & + p_{s,t}^{\text{dis}} - p_{s,t}^{\text{ch}} \leq \overline{P}_s \quad &&\forall s,t  \label{eq:reserve2} 
\end{align}

\subsubsection{Changes to the Real-Time Unit Commitment Model}
The RT-UC solves for the least cost commitment and dispatch, formulated as an MILP that co-optimizes energy and reserves over a two-hour horizon with a 15-minute resolution. The RT-UC can determine startup decisions only for short lead-time generators; i.e., natural gas and oil combustion turbine (CT) and natural gas and diesel internal combustion (IC). The commitment decisions for the first interval of the horizon are binding, while subsequent intervals are considered advisory and may change in future executions. 

To incorporate ESRs in the RT-UC framework, all modifications and new constraints related to ESRs in the DA model, \eqref{eq:obj_DA} - \eqref{eq:reserve2}, are replicated within the RT-UC formulation. Discharging, $p_{s,t}^{dis}$, and charging, $p_{s,t}^{ch}$, are scaled by the relative time resolution of RT-UC (i.e., by 1/4 for 15-minute intervals) to ensure power-to-energy consistency.

Because RT-UC is executed in a rolling horizon, the initial SoC for each execution is constrained in equation \eqref{eq:RTUC_initial} to the energy level from the preceding RT-ED solution, $e_{s,t}^{RTED}$ (i.e., the current system state). For the Day-Ahead Strategy, Equation \eqref{eq:RTUC_terminal} constrains the SoC in each interval  to the downscaled (15 min.) DA target. The other SoC strategies are described in Section \ref{sec:strategies}. Equations \eqref{eq:RTUC_initial} - \eqref{eq:RTUC_terminal} replace  \eqref{eq: initial_cdn}:
\begin{align}
    e_{s,t} &= e_{s,t}^{RTED} \quad \quad  &\forall s,t = t_0    \label{eq:RTUC_initial}\\
     e_{s,t}  & =   e_{s,t}^{DA}  \quad \quad     &\forall s,t > t_0  \label{eq:RTUC_terminal}
\end{align}
\subsubsection{Real-Time Economic Dispatch (RT-ED) Model}
The RT-ED model represents the Real-Time Market, and is formulated as a linear programming (LP) model that co-optimizes energy and reserves for a single 5-minute interval. A sequence of RT-ED executions are performed after each RT-UC execution, for the RT-ED intervals that correspond to the binding first interval of the RT-UC.  The RT-ED constraints are similar to those in the DA and RT-UC formulations. The exceptions are that commitment states are fixed for all resources to the most recent RT-UC solutions and are treated as exogenous parameters rather than optimization variables. All modifications to the DA and RT-UC related to ESRs are also applied to the RT-ED, interpolated to 5-minute resolution. 

\subsubsection{Real-Time Pricing (RT-PR) Model}
RT LMPs are calculated after the dispatch using a separate pricing engine with several relaxations rather than using the shadow prices from the RT-ED solution. In PJM, the pricing engine implements Fast-Start Pricing~\cite{guan2025pjm}, under which prices may reflect the startup costs of eligible fast-start units. The EMSF pricing engine simulates this approach. All modifications made to RT-ED to incorporate ESRs are also applied to RT-PR. The RT-PR is formulated as an LP in which the binary commitment states from the RT-UC solution are relaxed to be continuous on $[0,1]$, only for ``eligible`` fast-start units. For each dispatch interval, RT-ED is solved to obtain the dispatch quantities. Then RT-PR is solved for the same interval to determine prices. For full implementation details, see~\cite{EMSF_2025,guan2025pjm}.

\subsubsection{Out-of-Market Representation}
System operators manually commit generators or take other high-cost actions to ensure reliability. The EMSF represents operator-initiated manual, or ``out-of-market'' (OOM), actions with several blocks of high-cost, fast-start natural gas CTs, with a minimum output of zero, and no binding ramp rates. OOM units are not available in the DA or RT-UC.  OOM units are only dispatched in RT-ED when no other generation can meet the system's energy needs. OOM generation in RT-ED solutions is fixed in RT-PR, but these units cannot set prices. The supply of OOM consists of 200 MW blocks available in RT-ED at increasing marginal costs ranging from \$700 to \$1800 per MWh.

\subsection{Real-Time SoC Management Strategy Formulations}\label{sec:strategies}
When real-time markets are implemented via rolling horizon sequences of RT-UC and RT-ED with horizons that do not extend to the end of the day, an assumption for the SoC terminal condition is needed. In cost-minimizing engines, in the absence of a terminal SoC condition, there is no incentive to conserve any SoC beyond the current horizon. SoC management strategies differ in terms of how these terminal conditions are determined and enforced. In this study, we compare three alternative strategies for real-time SoC management:
\begin{itemize}
\item \textbf{Day-Ahead Strategy}: Enforces the DA SoC schedule with no flexibility in real-time.
\item \textbf{Administrative Pricing Strategy}: Dynamically adjusts real-time SoC requirements as conditions change, and allows SoC shortage with an administratively set penalty.
\item \textbf{Dynamic Pricing Strategy}: Dynamically adjusts real-time SoC requirements as conditions change, and adjusts the SoC shortage price based on scarcity pricing.
\end{itemize}

\subsubsection{Day-Ahead Strategy}
The formulation in Section \ref{sec:formulation} corresponds to the Day-Ahead strategy.  This formulation is consistent with studies that only solve a single intertemporal optimization over an entire day and reflects current practice in several markets that do not allow ESRs to alter their bids in real-time.
The DA model solves for a schedule of hourly SoC targets that satisfy the terminal conditions. The RT engines then enforce the downscaled DA SoC schedule without adjustments in real-time to account for load forecast changes. 

\subsubsection{Administrative Pricing Strategy} \label{sec:adminprice}
The Administrative Pricing strategy adjusts the DA SoC schedule in real-time in response to updated load forecasts. As in the Day-Ahead strategy, the DA model solves for a schedule of hourly SoC targets that satisfy the end-of-day terminal conditions. In real-time, each RT-UC execution re-optimizes the SoC targets within its horizon based on updated net load forecasts and the observed system state. The existing multi-interval RT-UC is used to co-optimize the modified SoC schedule with fast-start thermal commitment decisions. Both RT-UC and RT-ED allow violations (shortage) of the SoC constraint and are penalized in the objective function.  For this strategy, the penalty for SoC shortages is an administratively set, constant value, consistent with PJM's existing reserve products \cite{pjm2024shortage}. 

The RT-UC formulation is modified so that SoC is only constrained in the initial and terminal intervals. Equation \eqref{eq:RTUC_terminal} is replaced with \eqref{eq:RTUC_terminal_Modified}, which constrains SoC in the last interval of the horizon to be no lower than the DA target, $e_{s,t}^{DA}$. The terminal SoC constraint allows a shortage $\psi_{s,t}^{ue}$, which is penalized by $\pi^{ue}$ in the objective function \eqref{eq:obj_DA}.
\begin{align}
    e_{s,t} + \psi_{s,t}^{ue} & \geq e_{s,t}^{DA}  \quad \quad \forall s, t = t_{|T|} \label{eq:RTUC_terminal_Modified}
\end{align}   

 The single-interval RT-ED takes its SoC target from the RT-UC solution for binding interval, $(e_{s,t}^{RTUC})$, downscaled to a 5-minute resolution. The initial condition \eqref{eq:RTUC_initial} remains unchanged. Equation \eqref{eq:RTUC_terminal} is replaced with \eqref{eq:RTED_terminal_modified}: 
\begin{align}
    e_{s,t} + \psi_{s,t}^{ue} & \geq  e_{s,t}^{RTUC}  \quad \quad  \forall s, t = t_{|T|}   \label{eq:RTED_terminal_modified} 
\end{align}
As in the RT-UC, the RT-ED allows the SoC constraint to be violated. SoC shortages, $\psi_{s,t}^{ue}$, are penalized in the objective function using the pre-specified penalty,  $\pi^{ue}$. 

\subsubsection{Dynamic Pricing Strategy}
\label{sec:ShortagePricing} 
\begin{figure*}[b]
    \centering  
    \includegraphics[width=\linewidth]{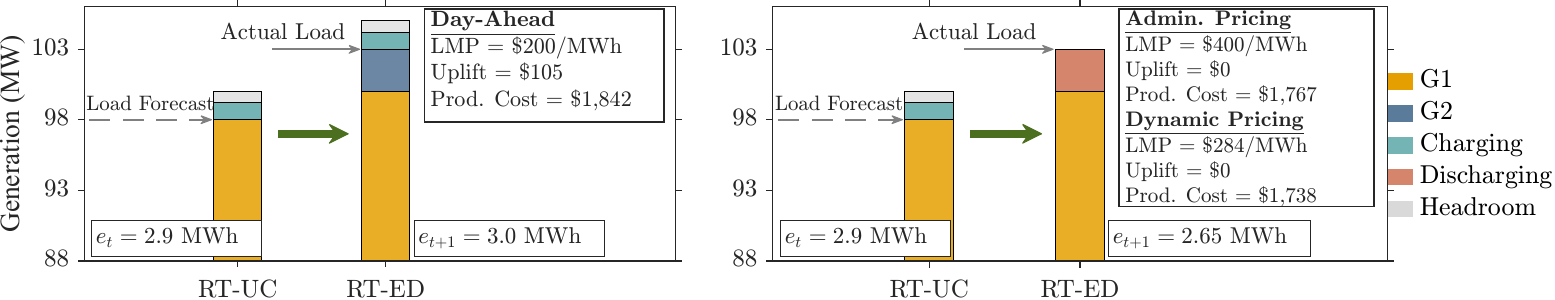}
    \caption{Illustrative Example: (left) Day-Ahead Strategy; (right) Administrative Pricing and Dynamic Pricing Strategies.}

    \label{fig:SimpleExample}
 \end{figure*}
The Dynamic Pricing strategy also adjusts the real-time SoC requirements to reflect the latest system conditions and forecasts, with the same formulation changes as shown above (Section \ref{sec:adminprice}).  In addition, this strategy dynamically adjusts the SoC shortage price, $\pi^{ue}_t$, to reflect the marginal value of the stored energy at each interval.

The SoC requirement should be violated only if the additional energy from ESRs is needed to avoid scarcity (i.e., load shedding). This strategy sets the SoC shortage price to the marginal value of responsive reserves. Hogan \cite{hogan2013electricity} defines the marginal value of responsive reserves $\pi_t(r)$ as 
\begin{align}
    \pi_t (r) = \rho_t (r) \bigl( \text{VoLL} - \lambda_t \bigr),
    \label{eq:reserve_price}
\end{align}
where $r$ is the quantity of available reserves, $\rho_t(r)$ is the probability that the realization of uncertain net load will exceed the available ramping capability $r$, VoLL is the administratively defined value of lost load, and $\lambda_t$ is the competitive energy price under non-scarcity conditions. As in \cite{hogan2013electricity}, we assume here that 1 MW of additional reserve provides 1 MW of capability if needed. Mays \cite{mays2021quasi} extended equation \eqref{eq:reserve_price} to a stochastic setting by defining $\lambda_t$ as the expected price conditional on non-scarcity conditions. The expression $(\text{VoLL} - \lambda_t)$ represents the net value of avoiding unserved load. Because values typically used for VoLL are several orders of magnitude above average price levels, this is often approximated as equal to the VoLL.

We assume that the available responsive reserves $r$ includes all energy that could be provided above the scheduled generation from committed thermal units and ESRs that can respond within the specified delivery time.  ESRs can contribute the difference, if any, between the scheduled discharge and the end-of-interval SoC target. Because $r$ in this context is distinct from the cleared reserve products (Section \ref{sec:formulation}), we replace $r$ in (\ref{eq:reserve_price}) with $H_t$ to represent the available headroom at time $t$. 

To determine the scarcity price, the probability $\rho_t$ of unserved load must be calculated. We define $\epsilon$ as the uncertain realization of net load above the expected net load (i.e., forecast error), with probability density $f_\epsilon(\cdot)$ and cumulative distribution function (CDF) $F_\epsilon(\cdot)$. The probability of scarcity is then
\begin{equation}
    \rho(H) = \mathbb{P}(\epsilon > H) = 1 - F_\epsilon(H) = 1 - \int_{-\infty}^{H} f_\epsilon(x)\,\mathrm{d}x.
\end{equation}
The distribution of load uncertainty $F_\epsilon$ can take any form.  For simplicity, here we assume a Gaussian approximation, $\epsilon \sim \mathcal{N}(0,\sigma^2)$, where $\sigma$ is estimated from historical forecast errors. The probability of shortage given headroom $H_t$ is 
\begin{equation}
    \rho_t(H_t) = \mathbb{P}\!\left(\epsilon > H_t \right)
    = 1 - \Phi\!\left(\frac{H_t}{\sigma}\right),
    \label{eq:scarcity_prob}
\end{equation}
where $\Phi(\cdot)$ is the standard normal CDF. The inverse CDF of other distributions can be substituted for $\Phi(\cdot)$.\\
\indent In the Dynamic Pricing strategy, the headroom and the corresponding SoC shortage price \eqref{eq:reserve_price} are recomputed before each RT-UC and again after each RT-UC and before the next RT-ED and RT-PR. The calculations use the time resolution of the RT-UC, e.g., $\Delta t = 15$ minutes. \\
\indent The headroom from thermal generators, $H_{g,t}$, is the upward ramping capability above the expected dispatch, $\hat{P}_{g,t}$:
\begin{equation}
    H_t^G
    =
    \textstyle\sum_{g\in\mathcal G}u_{g,t}^{*}
    \Bigl[
    \min\bigl(P_g^{\max},\,P_{g,t_0}^{\mathrm{ED}}+R_g\Delta t\bigr)
    -\hat{P}_{g,t}
    \Bigr]^+
    \label{eq:headroom_uc}
\end{equation}
where $[x]^{+} \equiv \max(x,0)$, $u_{g,t}^{*}\in\{0,1\}$ is the optimal commitment status of thermal generator $g$ from the most recent RT-UC solution, $P^{\text{ED}}_{g,t_0}$ is the initial condition for the RT-UC problem, and $R_g$ is the ramp rate of generator $g$. \\
\indent The headroom from energy storage $H_t^S$ represents the sum over all ESRs of the discharge capability of each unit above its scheduled output, constrained by its power rating and by the energy available above its DA SoC target:
\begin{equation}
\resizebox{0.89\linewidth}{!}{$
    \displaystyle
    H_t^S
    =
    \textstyle\sum_{s\in\mathcal S}
    \Bigl[
    \min\!\left(
    \frac{\eta_s}{\Delta t}
    \left[
    e_{s,t+\mathcal H}^{\mathrm{RT}^{*}}
    -
    e_{s,t+\mathcal H}^{\mathrm{DA}^{*}}
    \right]^+,
    \;
    \overline P_s^{\mathrm{dis}}
    -
    P_{s,t}^{\mathrm{dis}^{*}}
    \right)
    \Bigr]
    $}
    \label{eq:headroom_esr}
\end{equation}
where $e^{\text{RT}^{*}}_{s,t+\mathcal{H}}$ and $e^{\text{DA}^{*}}_{s,t+\mathcal{H}}$ denote the real-time and day-ahead SoC targets, respectively, at the end of the lookahead horizon $\mathcal{H}$, $\eta_s$ is the discharge efficiency of storage unit $s$, $\overline{P}^{\text{dis}}_s$ is the maximum discharge power limit of the ESR, and $P^{\text{dis}}_{s,t}$ is the scheduled discharge from the most recent RT-UC solution. The calculation of \eqref{eq:headroom_esr} for RT-UC and RT-ED use the appropriate time resolution and lookahead horizon length.\\
\indent The headroom $H_t$ used to calculate the SoC shortage price measures the total upward ramping capability from all resource types. For RT-UC, the headroom $H_t^\text{UC}$ is:
\begin{equation}
    H^{\text{UC}}_t =
    H_{t}^{\text{G,UC}} 
    + H_{t}^{\text{S,UC}} 
    \label{eq:headroom_uc}
\end{equation}
The headroom for RT-ED also includes the available out-of-market actions, $H_t^{\text{OOM}}$ that could be taken by the operator if the system has limited ramping capability:
\begin{equation}
    H^{\text{ED}}_t =
    H^{\text{G,ED}}_{t} 
    + H^{\text{S,ED}}_{t} 
    + H^{\text{OOM}}_t
    \label{eq:headroom_ed}
\end{equation}
The resulting SoC shortage price, $\pi^{\text{ue}}_t$, is effectively a dynamic 15-minute ramp price. 

\section{Illustrative Example} \label{sec:ResultsIllustrative}
To illustrate differences among real-time SoC management strategies, consider a simple system (Fig. \ref{fig:SimpleExample}) with thermal generators G$_1$ (100 MW, marginal cost $\$200$),  G$_2$ (5 MW, marginal cost $\$500$), and energy storage (10 MW / 10 MWh).  The initial SoC is 2.9 MWh and the next SoC target is 3.0 MWh. G$_1$ is committed and can meet the RT-UC forecast load of 98 MW. In the RT-ED, the actual load is 103 MW, so G$_1$ is not sufficient. The operator must either use G$_2$ or use stored energy. We compare the results from the three SoC management strategies below.

\subsubsection{Day-Ahead Strategy}
Under the Day-Ahead strategy (Fig. \ref{fig:SimpleExample}, left), the RT-ED strictly enforces the SoC target of $3.0$~MWh. The ESR must charge 0.1~MWh over the 5-minute interval by drawing 1.2 MW from the grid. The realized net load of 103~MW plus the 1.2 MW of charging creates a shortfall of 4.2~MW, and G$_2$ is committed out-of-market. Because OOM units cannot set the price, the price is set by the committed unit at \$200/MWh. Total cost is \$1,842 and G$_2$ receives a make-whole payment of $\$105$. This illustrates the issue that Potomac identified in ERCOT \cite{potomac2025ercot}. 

\subsubsection{Administrative Pricing Strategy}  Assume the Administrative Pricing strategy sets the penalty for SoC shortage to $\pi^{\text{ue}} = \$ 400$. Because this is less than the cost of G$_2$, the ESR will discharge the additional 3 MW to meet load (Fig. \ref{fig:SimpleExample}, right). The deviation from the original SoC target is able to meet load at lower cost $\$1,767$, and avoids the uplift payment. The energy price is set by the \$400/MWh penalty, which may not reflect the system's true opportunity cost. 

\subsubsection{Dynamic Pricing Strategy} The total headroom available in RT-ED is the sum of the thermal headroom above the RT-UC dispatch and G$_2$'s capacity (the ESR headroom is 0): 
\begin{equation*}
    H_t^{\text{tot}}
    = H_t^{\text{therm}} + H_t^{\text{OOM}}
    = \left(\overline{P}_g - \hat{D}^{\text{UC}}_t\right)
    + \overline{P}^{\text{OOM}}
    = 7~\text{MW}.
\end{equation*}

Using \eqref{eq:scarcity_prob} and a historical forecast error of $\sigma = 3.7$:
\begin{equation*}
\rho_t(H_t)
= 1 - \Phi\!\left(\frac{H_t^{\text{tot}}}{\sigma}\right)
= 1 - \Phi\!\left(\frac{7}{3.7}\right)
= 0.029.
\end{equation*}

This probability represents the \textit{ex ante} assessment of real-time ramp scarcity. By normalizing available system headroom against the historical forecast error, it quantifies the need for additional flexibility beyond the capacity scheduled in RT-UC. The resulting shortage price $\pi_t^{\text{ue}}$ is calculated as:
\begin{equation*}
\begin{aligned}
    \pi_t^{\text{ue}} &= \rho_t \bigl(\mathrm{VoLL}-\lambda_t\bigr)
          = \rho_t \bigl(\mathrm{VoLL}-C_g\bigr) \\
          &= 0.029\,(10{,}000-200)=\$284.20/\mathrm{MWh}.
\end{aligned}
\end{equation*}
Note that $\lambda_t = \$200$ because G$_2$ is the marginal unit that set the price in the RT-UC solution. Load is served at an even lower cost, $\$1,738$, with no uplift payment. Here, the resulting energy price is \$284.20/MWh, which reflects the opportunity cost of additional ramp. Production costs\footnote{A sensitivity analysis of the impact of SoC  price on reliability and cost for the PJM simulations was presented at IEEE PES General Meeting 2025 \cite{Wasti2025PESGM}.} are lower than under the Administrative Pricing strategy.

\section{Results from PJM Simulation} \label{sec:ResultsPJM}
This section presents the results from simulation of PJM's markets. The data and assumptions for the PJM case are provided first, followed by detailed results from one example day, and then the annual metrics from the simulated year.

\subsection{Case Study Assumptions} \label{sec:FutureScenario}
The PJM case study is calibrated with historical data from 2023. We simulate a future scenario developed by PJM \cite{pjm2023energytransition} that assumes, relative to 2023, a 40\% reduction in coal capacity, an 8\% reduction in natural gas capacity, a 500\% increase in wind, and a 300\% increase in solar capacity. Generator costs are derived from generator data in the S\&P Global Market Intelligence Platform \cite{sandp2026capitaliq}. The case study assumes 40 ESR units with an aggregate 12 GW of power and 56 GWh of energy. The ESRs consist of eight units of 500 MW / 3000 MWh (6-hour duration) and 32 units of 250 MW / 1000 MWh (4-hour duration). We assume that all ESRs have a 0.96 efficiency, operate between 20\% and 80\% of capacity, and have a variable O\&M cost of \$3/MWh. 

The EMSF replicates PJM's rolling-horizon structure: the DA model solves a 32-hour horizon in hourly steps; RT-UC solves a 2-hour horizon of 15-minute intervals (96 executions per day); and RT-ED and RT-PR solve for single 5-minute intervals (288 executions per day). The day-ahead net load for each simulated day is constructed from the historical 5-minute load, wind, and solar profiles for that date in \(2023\).  The profiles are then scaled for the future scenario: gross load is scaled by monthly load growth factors \cite{pjm2024loadreport}, subtracting renewable generation scaled by the projected capacity growth \cite{pjm2024RTEP}. Nuclear and hydro generation are also subtracted from gross load, with their capacities fixed at 2023 levels. The modified profiles are used to construct the day-ahead net load forecast for the simulated day.  Real-time net load forecasts add simulated forecast errors as the rolling horizon progresses. 

Forecast errors are sampled from distributions calibrated to historically observed forecast errors. Net-load forecast errors for RT-UC and RT-ED are simulated using an autoregressive (AR1) process, calibrated to 2022–23 wind, solar, and load forecast and realized values, then scaled using the same capacity growth factors applied to the profiles. From the 2022--23 data, the net load forecast error standard deviation is 721 MW (60-min ahead) and 494 MW (30-min ahead), and scales to 2,071 MW and 1,577 MW, respectively, for the future capacity mix. Simulated forecast errors reproduce the observed persistence of error direction. Each day is randomly assigned a forecast bias regime (under forecast, over forecast, or normal), and stochastic shocks of longer duration are periodically superimposed to represent sudden renewable or outage events. Each sample day consists of a realized net load profile for the target day and a sequence of 288 sample paths of forecast errors, one for each 5-minute RT market clearing. Simulations were performed on a Linux cluster (2.8 GHz 20-core Intel Xeon, 256 GB RAM) using GAMS 39.1.0 with CPLEX 12.10.0; MATLAB is used for data processing and information flow.

All SoC management strategies use a common DA solution for each simulated day. The Day-Ahead strategy downscales the hourly DA SoC schedule to a 5-minute resolution, with no deviation allowed. The Administrative Pricing strategy uses RT-UC to reallocate the SoC schedule within its two-hour horizon, constrained to meet the DA target at the end of the horizon.  SoC shortages are allowed but penalized at \$1400/MWh, a value just below the penalty for operating reserve shortages but more than the assumed OOM costs. The RT-ED SoC target is downscaled from the RT-UC SoC for the binding interval, and any shortage is penalized at \$1400/MWh. The Dynamic Pricing strategy is implemented with the same SoC target reoptimization as the Administrative Pricing strategy, but calculates the \textit{ex ante} SoC shortage price before every RT-UC and RT-ED executions (see Section \ref{sec:methodology}). For RT-UC, the shortage probability $\rho_t$ is calculated using a standard deviation of $\sigma = 2{,}000$ MW, based on the scaled 60-minute-ahead net-load forecast error, and assumes a lower bound of $\rho_{min} = 0.05$. For RT-ED, the assumed standard deviation accounts for both net load forecast error and generator forced outages.  Based on historical forced outages in PJM and 30-min ahead forecast errors, the shortage probability is calculated using $\sigma = 2{,}100$ MW and a lower bound of $\rho_{min} = 0.005$. 

\subsection{Results for Example Day} \label{sec:PJMExample}
\begin{figure}[H]
    \centering
   \includegraphics[width=0.97\linewidth]{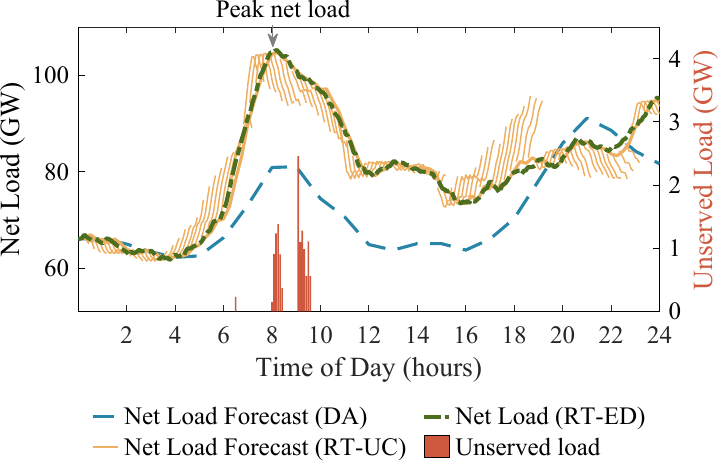}
    \caption{Net load forecasts (DA, RT-UC) and realized RT-ED net load (left y-axis), and unserved load under the DA Strategy (right y-axis) for example day (March 13).}
    \label{fig:NetLoad_NSE}
\end{figure}

To illustrate the main ideas, we compare the results from the Day-Ahead strategy and the Dynamic Pricing strategy for one example day, based on March 13 profiles (Fig. \ref{fig:NetLoad_NSE}). For this day, DA and RT-UC net load forecast errors, relative to the realized RT-ED net load, lead to periods of ramp scarcity. Under the DA\\ 
\noindent strategy, 1.1 GWh of unserved load occurs between 7 am  and 11 am. The unserved load is the result of day-ahead forecast errors combined with the lack of flexibility to adjust the SoC targets in real-time. Constraining ESRs to the DA SoC schedule in RT limits the ability of energy storage to address the ramp capability shortage. In this case, ESRs do not provide additional discharging during the ramp scarcity event, despite having sufficient stored energy.

\begin{figure}
    \centering
    \includegraphics[width=0.9\linewidth]{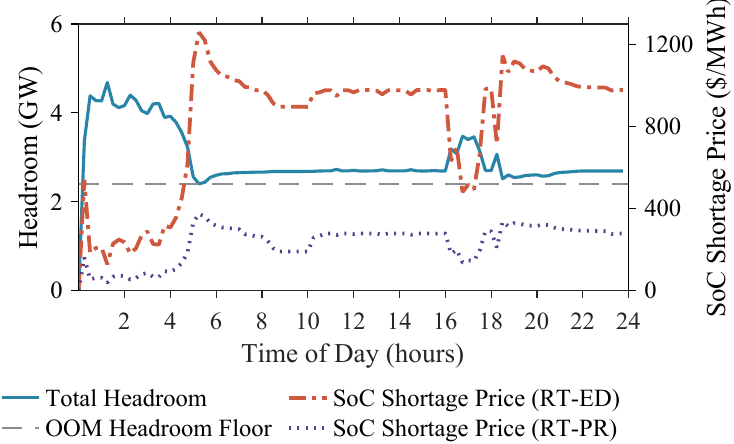}
    \caption{Total headroom and SoC shortage price under Dynamic Pricing Strategy for example day (March 13).}
    \label{fig:Headroom_73}
\end{figure}

Under the Dynamic Pricing strategy, the SoC shortage price evolves over the day as the system headroom within RT-ED changes (Fig.~\ref{fig:Headroom_73}). The OOM capacity establishes a lower bound on headroom with a corresponding upper bound on the scarcity probability. We assume a VoLL of \$10,000/MWh in RT-UC and RT-ED and \$3,000/MWh in RT-PR based on current PJM procedures. As the headroom declines during the morning ramp, the SoC shortage price increases, reflecting the increased marginal value of reserves. 

 \begin{figure}
    \centering
    \includegraphics[width=0.9\linewidth]{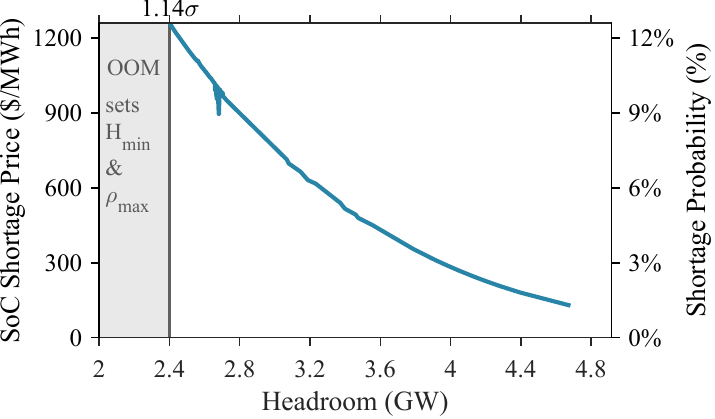}
    \caption{Demand curve for ramping reserve under Dynamic Pricing Strategy for example day (March 13).}
    \label{fig:ORDC_73}
\end{figure}

\begin{figure*}                    
\includegraphics[width=0.9\linewidth]{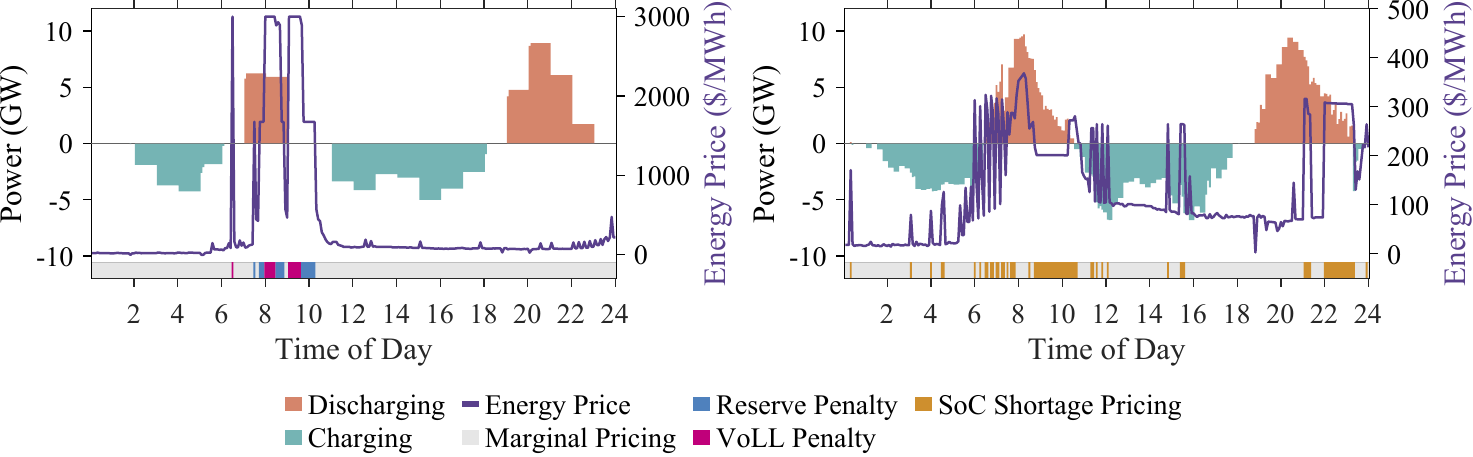}
        \caption{Comparison of ESR operation and energy prices under Day-Ahead (left) and Dynamic Pricing (right) strategies of March 13. Note the different secondary y-axis price scales. Bottom strip colors indicate price-setting unit types.}
    \label{fig:EnergyPriceComparison}
\end{figure*}

Using the results from all 288 5-min. intervals of this day, we construct the implicit endogenous demand curve for ramping reserve. Fig. \ref{fig:ORDC_73} shows the relationship between the available headroom $H$ and the corresponding shortage price $\pi$ (left axis) and the shortage probability $\Phi(-z)$ (right axis). The shaded region $(z\leq1.14\sigma)$ represents headroom equal to the available OOM capacity, which forms a lower bound on the shortage probability at $\Phi(-1.14) \approx 12.7 \%$. As the system headroom increases, the scarcity risk decreases non-linearly. The localized dip in the shortage price near $H = 2.7$ indicates those intervals in which RT-UC committed fast-start units. The fast-start commitments increase the energy price $\lambda_{t}$ and increase the headroom (more thermal capacity), which reduces the SoC price (Eq. \eqref{eq:reserve_price}.

The differences between the Day-Ahead and Dynamic Pricing strategies are illuminated by comparing the resulting charging/discharging behavior and energy prices under each strategy (Fig.~\ref{fig:EnergyPriceComparison}). The charging/discharging under the Day-Ahead strategy follows the DA schedule, and does not provide additional energy during the ramp scarcity event (8 - 10 am). The Dynamic Pricing strategy deviates from the DA schedule to charge more and earlier in advance of the event and discharge more during the event.  The impact on energy prices is more striking. The bottom bar color in Fig.~\ref{fig:EnergyPriceComparison} indicates whether the energy price was set by the marginal generator, the reserve shortage penalty, the VoLL, or the SoC shortage price. Under the Day-Ahead strategy, 84\% of prices are set by the marginal generator (below \$280/MWh) and 12\% of prices are set by the VoLL or reserve shortage penalties (above \$1500/MWh). Under the Dynamic Pricing strategy, 70\% of the energy prices are set by the marginal generator and 23\% of the prices are set by the dynamic SoC shortage price; none are set by reserve or load shedding penalties. The distribution of energy prices is more smoothly distributed by the Dynamic Pricing strategy.

\subsection{Annual Market Outcomes}
This section compares the outcomes from simulating a full year (365 days) of the PJM case study. We compare cost, reliability and price formation
across the three RT SoC management strategies. Table~\ref{tab:PercentileMetrics} summarizes the impacts by strategy on production cost and  unserved energy. The Administrative Pricing strategy eliminates all unserved load but achieves only a 3.8\% average reduction in production cost relative to the Day-Ahead strategy, due to a 4.4\% increase in OOM dispatch. The assumed SoC shortage penalty $\pi^{\text{ue}}=\$1400\text{/MWh}$ causes RT-ED to more frequently use OOM instead of stored energy to address ramp scarcity because the opportunity cost appears higher than is true for the system. The Dynamic Pricing strategy also eliminates all of the unserved energy, and reduces production costs by 12.1\% and OOM generation by 45\% relative to the Day-Ahead strategy. 

\begin{figure}
    \centering  
    \includegraphics[width=0.9\linewidth]{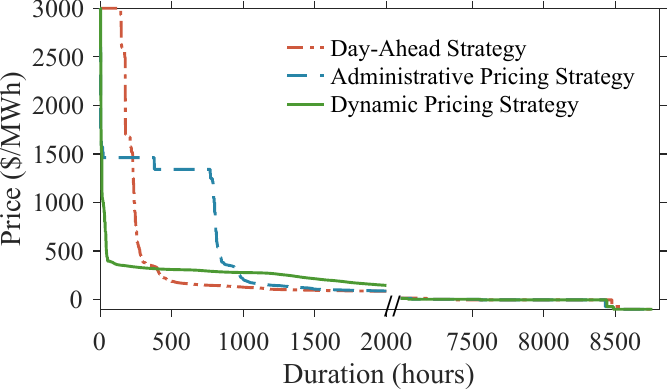}
    \caption{Energy price duration curve of three strategies.}
    \label{fig:RTPriceDuration}
 \end{figure}

Fig.~\ref{fig:RTPriceDuration} shows the real-time energy price duration curve for the three strategies. In the Day-Ahead strategy, the energy prices are set by the VoLL for 143 hours, above the reserve shortage penalty for roughly 230 hours, and less than \$200/MWh after 500 hours. Under the Administrative Pricing strategy, the energy price is set by either \$1{,}461/MWh (SoC shortage, discharging) or \$1{,}341/MWh (SoC shortage, charging) for nearly 800 hours. Not all of the hours with high energy prices were actually short of ramp capability; the constant SoC shortage penalty is too conservative at some of these times. The Dynamic Pricing strategy improves price formation by estimating the opportunity cost to the system based on current conditions \emph{before} a scarcity event occurs. The opportunity cost of holding stored energy is incorporated into energy prices. 

\begin{table}
    \centering
    \small
    \setlength{\tabcolsep}{4.0pt}
    \captionsetup{
        textfont={sc, footnotesize},
        labelfont=footnotesize,
        labelsep=newline
    }
    \caption{Distribution of outcomes over 365 days}
    \label{tab:PercentileMetrics}
    \begin{tabular}{lcccccc}
        \midrule
        \midrule
        \multicolumn{7}{c}{ Production Cost (Mil \$)} \\
        \midrule
        \textbf{Percentile} & \textbf{p05} & \textbf{p25} & \textbf{p50} & \textbf{p75} & \textbf{p99} & \textbf{Mean} \\
        \midrule
        Day-Ahead Strategy                & 24 & 32 & 39 & 47 & 75 & 41 \\
        Admin.  Pricing Strategy  & 24 & 31 & 38 & 45 & 70 & 39 \\
        Dynamic Pricing Strategy   & 20 & 28 & 35 & 42 & 66 & 36 \\
        \midrule
        \midrule
        \multicolumn{7}{c}{Unserved Energy (MWh)} \\
        \midrule
        \textbf{Percentile} & \textbf{p05} & \textbf{p25} & \textbf{p50} & \textbf{p75} & \textbf{p99} & \textbf{Mean} \\
        \midrule
        Day-Ahead Strategy                & 0 & 48  & 140 & 301 & 4{,}441 & 309 \\
        Admin.  Pricing Strategy & 0 & 0   & 0   & 0   & 0       & 0   \\
        Dynamic Pricing Strategy  & 0 & 0   & 0   & 0   & 0       & 0   \\
        \midrule
        \midrule
    \end{tabular}
\end{table}

\begin{figure}
    \centering  \includegraphics[width=0.9\linewidth]{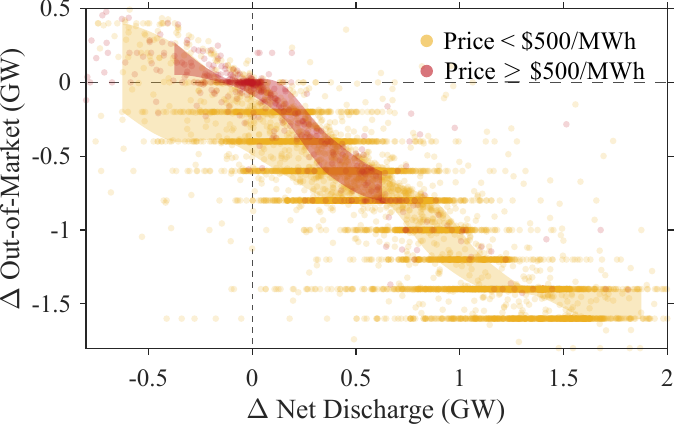}
   \caption{$\Delta\text{OOM}$ vs. $\Delta$ Net Discharge during Administrative Pricing scarcity periods ($>\$500$/MWh) from annual data; colors indicate Dynamic Pricing Strategy's energy price.}
    \label{fig:PdisOOMBin}
\end{figure}
Both the Administrative and the Dynamic Pricing strategies eliminate unserved energy because they adjust the SoC schedule using updated forecasts. The Administrative Pricing strategy results in higher costs than Dynamic Pricing because of the difference in how they respond to 30-min. ahead forecast errors (i.e., between RT-UC and RT-ED). The benefits of dynamically adjusting the shortage penalty is illustrated by the relationship between energy prices, OOM generation, and net discharging from energy storage. Fig. \ref{fig:PdisOOMBin} shows this relationship for all  5-min. intervals for which the energy price under Administrative Pricing exceeded \$500/MWh (i.e., SoC shortage conditions). The colors indicate whether the energy price under the Dynamic Pricing strategy is above or below \$500/MWh. When the energy price under Dynamic Pricing is lower (lighter colors), discharging from storage displaces OOM generation, relative to Administrative Pricing. Only 4.5\% of these intervals have high prices under Dynamic Pricing, suggesting that the administrative penalty is inflated relative to the true opportunity cost at those times. The horizontal clusters of intervals indicate the number of discrete 0.2 GW blocks of OOM dispatched. 

\section{Discussion}
\label{sec:Discussion}
We develop a coordinated SoC management mechanism that balances its use between intraday energy arbitrage and addressing short-term ramp scarcity. Errors in day-ahead forecasts relative to real-time are addressed by adjusting the DA SoC schedule using the existing RT-UC engine. Changes in forecasts between RT-UC and RT-ED are handled by allowing deviations from the SoC target in RT-ED. Lower production costs and improved price formation are achieved when SoC shortages are penalized using the marginal value of reserves in each interval.

The Dynamic Pricing strategy is motivated by the insight that SoC management in real-time is not a distinct problem from pricing operating reserves or flexible ramping reserves; these are all the same problem. Just as reserve prices compensate thermal generators for foregone energy revenue in the current interval, the SoC shortage price indicates the marginal value of conserving stored energy, which is balanced against the immediate reliability benefit of discharging now. Our approach is compatible with energy markets that have implemented flexibility products of other durations.  The marginal value of each product will vary with the delivery time, but should be agnostic to resource type. 

The adoption of explicit flexibility products by several U.S. wholesale markets indicates that energy prices alone are insufficient to procure adequate ramping capability, particularly as the share of zero marginal cost resources grow. As the scarce resource shifts from energy to ramping reserve, improved price formation will become essential to maintain system reliability and incentivize investment in flexible capacity. Our findings suggest that dynamic pricing of ramp scarcity can provide a market-based mechanism for valuing flexibility within markets. 

Our approach also highlights the importance of the often ignored real-time unit commitment engine, which is an integral component of market operations.  Although itself not a market, the greatest impacts from RT market design changes are often achieved by inducing a more efficient commitment pattern. For SoC management, outcomes can be improved by  co-optimizing the adjusted SoC schedule with thermal commitment decisions. 

\section{Acknowledgments}
The authors thank  Anirudh Subramanyam, Chiara Lo Prete, Ashish Radhakrishnan from the Pennsylvania State University, Yonghong Chen from National Laboratory of The Rockies and Uday V. Shanbhag from the University of Michigan for  helpful suggestions. We are  grateful to EPRI for the opportunity to present this work, and to the participants for feedback.

\bibliographystyle{IEEEtran}
{\footnotesize \bibliography{IEEEabrv,References,Storage_Zotero}}

\end{document}